\documentclass[twocolumn,superscriptaddress,nobibnotes,footinbib,amsmath,floatfix,aps,prapplied,citeautoscript,reprint]{revtex4-1}

\usepackage{graphicx}% Include figure files
\usepackage{dcolumn}% Align table columns on decimal point
\usepackage{bm}% bold math
\usepackage{amsmath,bm,multirow}
\usepackage{amsfonts}
\usepackage{float}
\usepackage{color}
\usepackage[draft]{changes} % final OR draft
\usepackage[normalem]{ulem}
\definechangesauthor[name={yi}, color=blue]{Yi}

\newcommand{\nw}{Department of Materials Science and Engineering, Northwestern University, Evanston, IL 60208, USA}

\newcommand{\yale}{Department of Applied Physics, Yale University, New Haven, CT 06520, USA}
\newcommand{\esi}{Energy Sciences Institute, Yale University, West Haven, CT 06516, USA}

\usepackage{float}
\usepackage{graphicx}
\usepackage{sidecap}
\usepackage{tabularx}
\usepackage{sidecap}
\newcommand{\etal}{\textit{et al}. }

\begin{document}
%\preprint{xxxx}
%\thanks{Footnote to title of article.}
\title{Impact of Temperature-Dependent Rattling Phonons on Lattice Dynamics and Thermal Transport in Ag$_{6}$Ge$_{10}$P$_{12}$}

\author{Yi Xia}
\email{yimaverickxia@gmail.com}
\affiliation{\nw}

\author{Vidvuds Ozoli\c{n}\v{s}}
%\email{vidvuds.ozolins@yale.edu}
\affiliation{\yale}
\affiliation{\esi}

\date{\today}

%--------------------------------------------------------------------------------------------------------------------------
% Abstract
%--------------------------------------------------------------------------------------------------------------------------
\begin{abstract}
Crystalline compounds exhibiting low-frequency rattling phonons constitute an important class of high-performance thermoelectrics owning to their intrinsically very low lattice thermal conductivity ($\kappa_{l}$). Theoretical approach that is capable of revealing the physical origin and accurately predicting $\kappa_{l}$ is of particular interest, which, however, still remains an outstanding challenge. In this study, we perform a case study of lattice dynamics and thermal transport properties of Ag$_{6}$Ge$_{10}$P$_{12}$, which has recently been identified as a high-performance thermoelectric phosphide due to low $\kappa_{l}$, arising from rattling vibrations associated with Ag$_{6}$ clusters. Analysis within a first-principles-based lattice-dynamics framework based on self-consistent phonon theory reveals a strong temperature dependence of rattling phonons due to high-order anharmonic interactions. Anharmonic hardening of the rattling optical modes has a strong effect on the lifetimes of heat-carrying acoustic phonons by decreasing the rate of three-phonon combination processes. This mechanism results in a significant increase in $\kappa_l$ and changes its temperature dependence to $1/T^{0.64}$.
\end{abstract}
%\pacs{Valid PACS appear here}%
%\keywords{Temperature effect, Thermoelectric, Lattice thermal conductivity, Anharmonicity, Phonon renormalization, Phonon-phonon interactions}
\maketitle

%------------------------------------------------------------------------
% Introduction
%------------------------------------------------------------------------
\section{\label{sec:introduction} Introduction}

Lattice thermal conductivity ($\kappa_{l}$) is a key parameter in determining properties of materials for applications such as thermoelectrics, thermal barrier coatings, and integrated circuits~\cite{Bell1457}. Thermoelectric materials require low $\kappa_{l}$ to maximize the energy conversion efficiency characterized by the figure of merit, $zT=S^2\sigma T/(\kappa_{e}+\kappa_{l})$, where $T$ is the absolute temperature, $S$ is the thermopower, $\sigma$ is the electrical conductivity, and $\kappa_{e}$ is the electronic thermal conductivity~\cite{rowe1995crc}. Various strategies for achieving glasslike ultralow $\kappa_{l}$ have been proposed, such as strong intrinsic anharmonicity induced by stereochemically active lone pair electrons~\cite{Skoug2011,Nielsen2013,Lu2013}, structural complexity~\cite{Brown2006}, and rattling phonons in guest-host systems~\cite{Cohn1999,Takabatake2014,Ren2017}.

\begin{figure}[htp]
	\includegraphics[width = 0.75\linewidth]{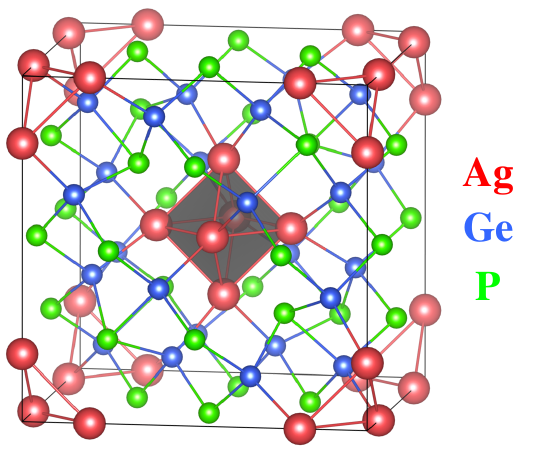}
	\caption{ 
	Crystal structure of Ag$_{6}$Ge$_{10}$P$_{12}$, wherein an octahedral cluster composed of six Ag atoms is highlighted. Ag, Ge and P atoms are colored in red, blue and green, respectively. Crystal structure is visualized using VESTA software~\cite{vesta2008}.
	}
	\label{fig:structure}
\end{figure}

Recently, Nuss \etal \cite{Nuss2017} and Shen \etal \cite{Shen2018} independently demonstrated a remarkably high $zT$ of approximately 0.6 in the mid-temperature range for a ternary transition metal phosphide, Ag$_{6}$Ge$_{10}$P$_{12}$. The high thermoelectric performance was attributed to the presence of multiply degenerate hole pockets with highly anisotropic (light and heavy) effective masses and intrinsically low $\kappa_{l} \approx$~1.0 W m$^{-1}$ K$^{-1}$ at $T=700$~K. The low $\kappa_{l}$ is related to the unique crystal structure of this compound~\cite{Nuss2017,Shen2018}, which belongs to the $\bar{I}43m$ space group with a complex primitive cell containing 28 atoms, as shown in Fig.~\ref{fig:structure}. The key feature of the structure is the presence of octahedral clusters composed of six Ag atoms in the center and corner positions of a body-centered cubic lattice. In contrast with other guest-host systems, such as clathrates and skutterudites where rattling phonons are due to guest atoms filling large voids in the host crystal~\cite{Cohn1999,Takabatake2014,Ren2017}, the localized rattling-like vibrations in Ag$_{6}$Ge$_{10}$P$_{12}$ are tied to Ag$_{6}$ clusters, which are weakly coupled to the surrounding network of covalent bonds formed by Ge and P atoms, exhibiting considerably larger atomic displacements than the ones from other atoms~\cite{Nuss2017}.

In view of the fact that thermoelectric performance of most covalently bonded phosphides is limited by high values of $\kappa_l$ arising from stiff interatomic bonds and light atomic masses, it is interesting and potentially useful to gain a comprehensive understanding of the lattice dynamics, anharmonicity, and low $\kappa_l$ in Ag$_6$Ge$_{10}$P$_{12}$. An earlier first-principles study~\cite{Shen2018} based on the framework of anharmonic lattice dynamics (ALD), first-order perturbation theory and linearized Peierls-Boltzmann transport equation (PBTE)~\cite{Maradudin1962,peierls1996quantum,wallace1998thermodynamics,broido,Esfarjani2011,Garg2011,McGaughey2019,LINDSAY2018106} revealed that the rattling phonons associated with Ag$_6$ clusters reduce $\kappa_{l}$ via cutting through the acoustic region, introducing ``avoided crossings'' and suppressing both phonon group velocities and lifetimes (see details in Ref.~[\onlinecite{Shen2018}]). These mechanisms were qualitatively similar to those suggested by previous first-principles studies on skutterudites~\cite{liwu2014,Wuli2015} and clathrates~\cite{tadano}. However, quantitative agreement between theoretically calculated and experimentally measured $\kappa_{l}$ was not achieved. For instance, the calculated $\kappa_{l} \approx 0.5$~W m$^{-1}$ K$^{-1}$ at $T=700$~K is lower by a factor of two than the experimental value of approximately 1.0 W m$^{-1}$ K$^{-1}$~\cite{Nuss2017,Shen2018}. It is interesting that similar discrepancies between experimental data and theoretical first-principles calculations are fairly common in the recent literature on ``rattling'' compounds such as skutterudites (YbFe$_{4}$Sb$_{12}$)~\cite{Wuli2015} and clathrates (Ba$_{8}$Ga$_{16}$Ge$_{30}$)~\cite{tadano}.

Tadano and Tsuneyuki recently reported that strong quartic anharmonicity of the Ba atoms in Ba$_{8}$Ga$_{16}$Ge$_{30}$ causes hardening of the frequencies of the rattling phonons with increasing temperature, which significantly affects $\kappa_{l}$~\cite{Tadano2018}. It remains unclear whether such a mechanism is universal and might explain discrepancies in $\kappa_l$ between theory and experiments in other systems with rattling phonons. To address these questions, we perform first-principles calculations of the lattice dynamics and thermal transport properties of Ag$_{6}$Ge$_{10}$P$_{12}$ using a self-consistent phonon (SCPH) scheme that includes anharmonic phonon renormalization (APRN) due to quartic anharmonicity and accounts for three-phonon interactions between renormalized phonons. We find that quartic anharmonicity shifts the rattling modes of the Ag$_6$ clusters to higher frequencies and reduces the scattering and suppression of group velocities of long-wavelength acoustic phonons. This leads to an improved agreement for $\kappa_l$ between theory and experiment both in magnitude and the observed temperature dependence.

%------------------------------------------------------------------------
% Method
%------------------------------------------------------------------------
\section{\label{sec:method} Method}

The conventional perturbative approach to anharmonic lattice dynamics starts from the harmonic phonon dispersion calculated from a dynamical matrix which only includes the second derivative of the Born-Oppenheimer potential energy surface (PES). Anharmonicity is then taken into account as a weak perturbation and expanded in diagrammatic series of which only the first few terms are amenable to practical calculation. It is common for such calculations to only account for three-phonon processes arising from the 3rd-order derivatives of the Born-Oppenheimer PES. In systems with strong anharmonicity and/or at high temperatures this procedure stops being qualitatively or quantitatively accurate. The situation is even more delicate in compounds with harmonically unstable phonons that are stabilized at high temperature by anharmonic contributions to PES. One effective way to overcome this theoretical challenge is by correcting phonon frequencies in either a perturbative or self-consistent manner, which, here we refer to anharmonic phonon renormalization (APRN). Recently, several first-principles-based APRN schemes~\cite{Souvatzis2009,Hellman2011,Errea2014,Tadano2015,Roekeghem2016,yixia2018} have been introduced, relying on either (i) extracting effectively renormalized interatomic force constants (IFCs) in real space~\cite{Hellman2011,Errea2014,Roekeghem2016} or (ii) directly estimating renormalized phonon frequencies in reciprocal space~\cite{Souvatzis2009,Tadano2015}. In this study, we employ the self-consistent phonon (SCPH) theory~\cite{Hooton1955,Koehler1966,Werthamer1970} formulated in the reciprocal space, which has been derived using either (i) a variational approach that minimizes the anharmonic free energy~\cite{Errea2011} or (ii) the many-body Green's function theory~\cite{Tadano2018}. The resultant SCPH equation that accounts for the first-order correction from quartic anharmonicity in the diagonal form is
\begin{equation}
\label{eq:SCPH1}
\Omega^2_{q} = \omega^2_{q}+2\Omega_{q}I_{q},
\end{equation}	
where $\omega_{q}$ is the harmonic phonon frequency associated with phonon mode indexed by $q$ (a composite index of both phonon branch and wave vector) and $\Omega_{q}$ is the renormalized phonon frequency. The quantity $I_{q}$ is defined as
\begin{equation}
\label{eq:SCPH2}
I_{q} = \frac{\hbar}{8} \sum_{q^{\prime}} \frac{\Phi^{(4)}(q,-q,q^{\prime},-q^{\prime})}{\Omega_{q}\Omega_{q^{\prime}}} \left[ 1+2n\left(\Omega_{q^{\prime}}\right) \right],
\end{equation}
where $\hbar$, $n$, and $\Phi^{(4)}$ are the reduced Planck constant, phonon population, and the reciprocal representation of the 4th-order IFCs~\cite{Tadano2015}, respectively. The temperature dependence of the SCPH equation is contained in the phonon population that follows the Bose-Einstein statistics. Since $\Omega_{q}$ depends on itself and $I_{q}$, the later of which in turn relies on $\Omega_{q}$, Eqs.~\eqref{eq:SCPH1} and \eqref{eq:SCPH2} are solved self-consistently. We implemented this scheme in the ShengBTE package~\cite{shengbte}. It is worth noting that a more advanced SCPH scheme including off-diagonal terms that enable updating phonon eigenvectors (polarization mixing) has been recently developed by Tadano and Tsuneyuki~\cite{Tadano2015}. To estimate the contributions from the off-diagonal terms, we performed additional APRN calculations using a recently developed real-space-based scheme which naturally incorporates polarization mixing~\cite{yixia2018,yixiagete2018}. The resulting renormalized phonon dispersion curves are in excellent agreement with those from SCPH calculations (see Fig.~\ref{fig:compaprn} in~Appendix \ref{sec:compaprn}), indicating that the effect of the off-diagonal terms is small. We also note that even at $T =$ 0~K zero-point correction due to the quantum nuclear effect may also lead to significant phonon frequency renormalization~\cite{Errea2011,Errea2015,Shulumba2017}. However, in this study we adopt a convention that phonon dispersions at $T = $ 0~K refers to the results obtained using regular harmonic approximation without considering the quantum nuclear effect.

\begin{figure*}[htp]
	\includegraphics[width = 1.0\linewidth]{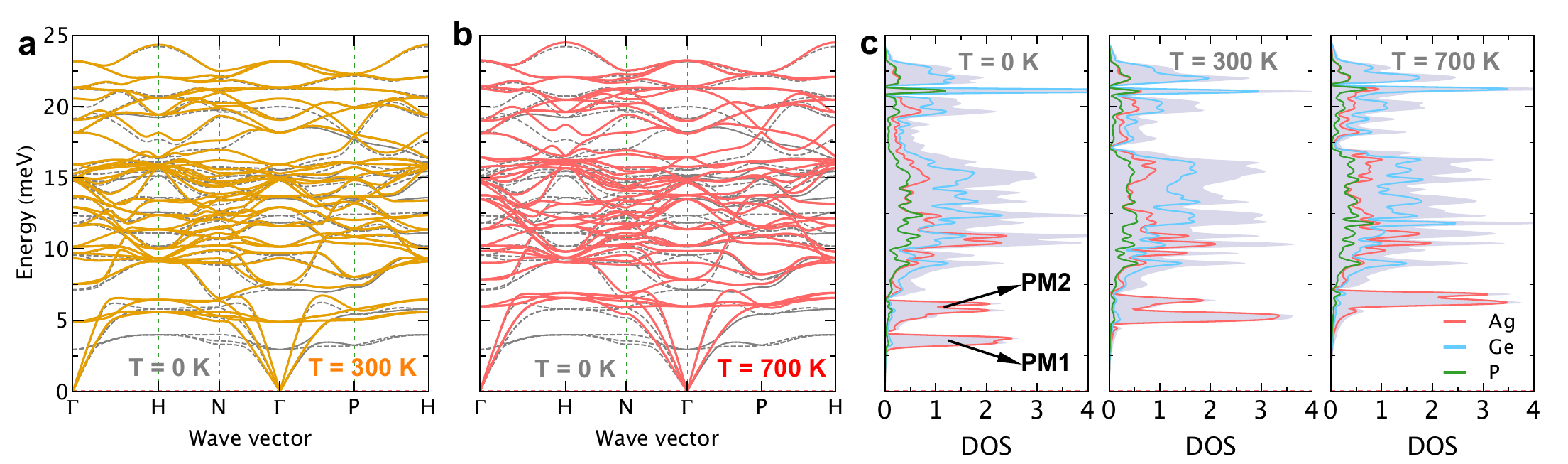}
	\caption{ 
	(a) Renormalized phonon dispersions curves at $T=300$~K (solid orange lines) in comparison with the phonon dispersion computed at $T=0$ K (dashed gray lines). (b) Same as (a) but at $T=700$~K. (c) Atom-projected phonon DOS at $T=0$, 300, and 700~K, respectively.
	}
	\label{fig:phonon-new}
\end{figure*}

%--------------------------------------------------------------------------------------------------------------------------
% Computational details
%--------------------------------------------------------------------------------------------------------------------------
%\section{\label{sec:details} Computational details}

We performed density-functional-theory (DFT)~\cite{dft} calculations using the Vienna {\it Ab\ Initio\/} Simulation Package (VASP)~\cite{Vasp1, Vasp2, Vasp3, Vasp4}. The projector-augmented wave (PAW)~\cite{paw} method was used in conjunction with the Perdew-Burke-Ernzerhof revised PBEsol~\cite{pbe,Perdew2008} version of the generalized gradient approximation (GGA)~\cite{gga} for the exchange-correlation functional~\cite{dft}. The basis set included plane waves with a kinetic energy cutoff value of 520 eV, and the Brillouin zone was sampled on a $\mathbf{k}$-point mesh with density equivalent to the $\Gamma$-centered meshes of 4$\times$4$\times$4 for the primitive cell. In both structural relaxation and self-consistent DFT calculations the force and energy convergence thresholds were set to $10^{-3}$ eV/\r{A} and $10^{-8}$ eV, respectively. Supercell (2$\times$2$\times$2) structures with thermalized atomic displacements at 300~K were used in compressive sensing lattice dynamics (CSLD)~\cite{csld,2018arXiv180508904Z} to extract both harmonic and anharmonic IFCs up to the 4th order. The fitted IFCs lead to a force prediction error of less than 5.0\% during an out-of-sample testing, indicating that a very good representation of the DFT PES was achieved.

We computed $\kappa_{l}$ using the linearized PBTE under the single mode relaxation time approximation (SMRTA)~\cite{peierls1996quantum},
\begin{equation}
\label{eq:PBTE}
\kappa_{l} = \frac{1}{NV} \sum_{q}  C_{q}   \mathbf{v}_{q}  	\otimes \mathbf{v}_{q} \tau_{q},
\end{equation}
where $N$, $V$, $C_{q}$, $\mathbf{v}_{q}$ and $\tau_{q}$ are the number of sampled phonon wave vectors, the volume of the primitive cell, mode-resolved heat capacity, group velocity, and lifetime, respectively. We estimated $\tau_{q}$ by accounting for both phonon-phonon and phonon-isotope scatterings~\cite{Tamura1983,Tamura1984}. The phonon-phonon scattering rates ($\tau_{3,q}^{-1}$) was computed considering three-phonon interactions~\cite{Maradudin1962,Cowley1968,srivastava1990physics,shengbte,Tianli2016} 
\begin{equation}
\label{eq:ThreeP}
\begin{split}
\tau_{3,q}^{-1} &= \sum_{q^{\prime}q^{\prime\prime}} \left\{ \frac{\left(1+n_{q^{\prime}}+n_{q^{\prime\prime}}\right)}{2}\zeta_{-} + \left(n_{q^{\prime}}-n_{q^{\prime\prime}}\right)\zeta_{+} \right\},
\end{split}
\end{equation}
with
\begin{equation}
\label{eq:ThreeL}
\begin{split}
\zeta_{\pm} & = \frac{\pi\hbar}{4N} \vert V_{\pm}^{(3)} \vert^{2} \Delta_{\pm} \frac{ \delta ( \Omega_{q}\pm\Omega_{q^{\prime}}-\Omega_{q^{\prime\prime}} ) }{ \Omega_{q}\Omega_{q^{\prime}}\Omega_{q^{\prime\prime}} },
\end{split}
\end{equation}
and 
\begin{equation}
\label{eq:ThreeV}
\begin{split}
V_{\pm}^{(3)}  = \sum_{ b,l_{1}b_{1},l_{2}b_{2} } \sum_{ \alpha\alpha_{1}\alpha_{2} } & \Phi_{ 0b,l_{1}b_{1},l_{2}b_{2} }^{ \alpha\alpha_{1}\alpha_{2} } \frac{ e_{b \alpha}^{q} e_{b_1 \alpha_1}^{\pm\ q^{\prime}} e_{b_2 \alpha_2}^{-q^{\prime\prime}} }{ \sqrt{ m_{b} m_{b_1} m_{b_2} } } \\
& \cdot e^{\pm i \mathbf{q}^{\prime} \cdot \mathbf{r}_{l_1} } e^{- i \mathbf{q}^{\prime\prime} \cdot \mathbf{r}_{l_2} },
\end{split}
\end{equation}	
where $l$, $b$, and $\alpha$ indicates the primitive cell, basis atom, and cartesian coordinate, respectively; $m$ is the atomic mass; $\mathbf{r}$ is the lattice vector of the primitive cell; $e^{q}$ is the phonon eigenvector; $\Phi_{ 0b,l_{1}b_{1},l_{2}b_{2} }^{ \alpha\alpha_{1}\alpha_{2} }$ is the real-space representation of the 3rd-order IFCs.  In the three-phonon processes, $\zeta_{-}$ represents the  splitting process ($q \rightarrow q^{\prime}+q^{\prime\prime}$), and $\zeta_{+}$ indicates the combination process ($q+q^{\prime} \rightarrow q^{\prime\prime}$), wherein momentum conservation is strictly enforced as indicated by $\Delta_{\pm}$ and energy conservation is enforced by $\delta$ functions, which are approximated by adaptive Gaussian smearing \cite{wuli2012}. To obtain the renormalized phonon frequency $\Omega_{q}$ at finite temperatures, we solved the SCPH equations using a phonon wave vector $\mathbf{q}$ mesh of 2$\times$ 2$\times$2 (equivalent to 224 atoms), the convergence of which was confirmed by comparing to a denser $\mathbf{q}$ mesh of 4$\times$ 4$\times$4 (equivalent to 1792 atoms). The renormalized phonon frequencies and eigenvectors were used to perform the inverse Fourier transform to obtain renormalized IFCs, which were later used to construct dynamical matrix at arbitrary $\mathbf{q}$ point. We solved PBTE using the ShengBTE package~\cite{shengbte,PhysRevB.85.195436}. Based on rigorous convergence tests, we find that  $\kappa_{l}$ is converged to sufficient accuracy (within 5\%) using a $\mathbf{q}$ mesh of 12$\times$12$\times$12 and including up to the 6th-nearest neighbors for the 3rd-order anharmonic interactions. We additionally confirmed that solving PBTE in an iterative manner~\cite{OMINI1995101,omini2} leads to negligible changes in $\kappa_{l}$.

%--------------------------------------------------------------------------------------------------------------------------
% Results and discussion
%--------------------------------------------------------------------------------------------------------------------------
\section{\label{sec:results} Results and discussion}

%--------------------------------------------------------------------------------------------------------------------------
\subsection{Phonon dispersions}

\begin{figure*}[htp]
	\includegraphics[width = 1.0\linewidth]{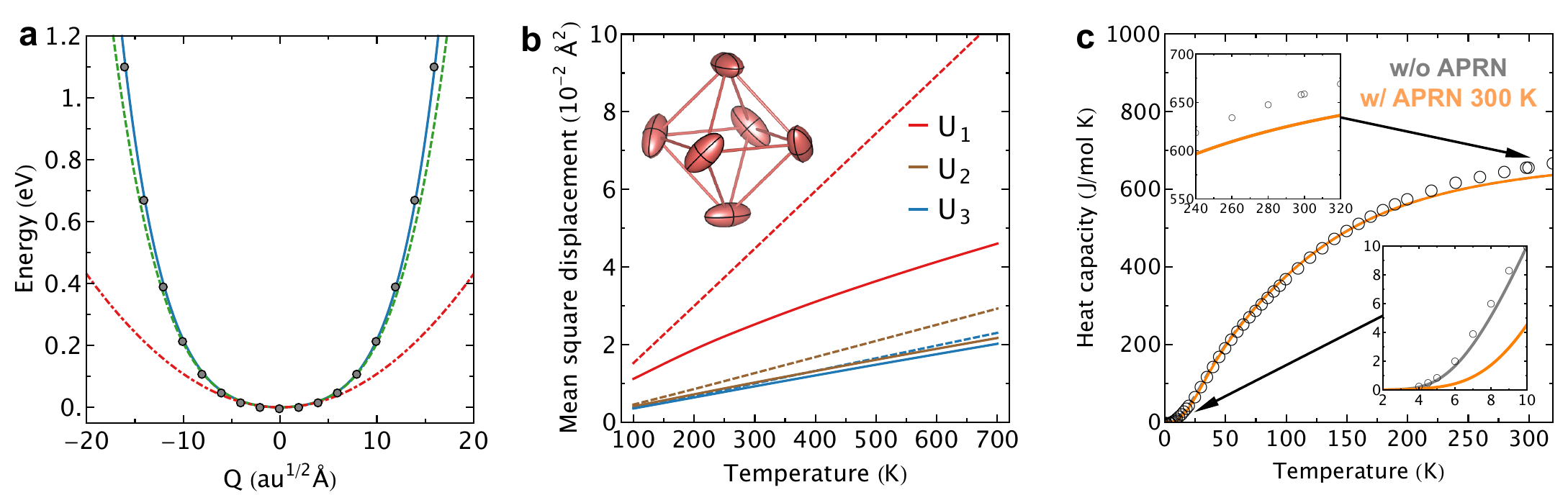}
	\caption{ 
	(a) DFT-calculated potential energy surface (solid disks) of the low-lying zone-center rattling mode as a function of the vibrational amplitude ($Q$) in the normal mode coordinate. The red, green and blue lines represent polynomial Taylor expansions around the equilibrium position up to 2nd, 4th and 6th order, respectively. (b) Calculated principle mean square atomic displacements ($U_1$, $U_2$, $U_3$) of Ag atoms as functions of temperature. The dashed and solid lines correspond to the results calculated without and with anharmonic phonon renormalization (APRN), respectively. The inset depicts a representative thermal displacement ellipsoids of Ag$_6$ cluster with the principal ellipses denoted by black solid lines, which is visualized using the VESTA software~\cite{vesta2008}. (c) Calculated constant-volume heat capacity (gray and orange lines) as a function of temperature in comparison with experimental measurements (empty circles)~\cite{Gmelin1989}. The gray and orange lines correspond to the results calculated using phonon dispersions at $T$= 0 K and $T$= 300 K, respectively. The insets highlight the comparison at the low- and high-temperature end, respectively.
	}
	\label{fig:aprnana}
\end{figure*}

Fig.~\ref{fig:phonon-new}(a) and (b) compare the computed APRN phonon dispersions at finite temperature to those at $T=0$~K without APRN. In line with our earlier study~\cite{Shen2018}, the critical feature of the $T=0$~K phonon dispersions is the presence of a set of optical phonon modes with very low frequencies (less than $5$ meV), cutting through the acoustic region and exhibiting rattling behavior. The corresponding atom-projected phonon density of states (DOS) in Fig.~\ref{fig:phonon-new}(c) reveals that these rattling phonon modes are associated with vibrations of Ag atoms and give rise to a sharp DOS peak centered at approximately 3.5~meV. An additional DOS peak dominated by Ag vibrations is centered at slightly higher frequencies of approximately 5.5~meV. We refer to the phonon modes associated with these two Ag DOS peaks as PM1 and PM2, respectively, as denoted in Fig.~\ref{fig:phonon-new}(c). Subsequent analysis will show that these modes have a large impact on $\kappa_{l}$. Upon performing APRN, the PM1 and PM2 modes display strong temperature dependence, while the acoustic and high-frequency optical modes corresponding to majority Ge and P vibrations remain largely unaffected. Specifically, a significant frequency shift of the zone-center PM1 modes from 2.9~meV to 4.9~meV is observed when the temperature increases to 300~K, and PM1 is further shifted to 5.9~meV at 700~K. Meanwhile, the PM2 modes also display phonon hardening at elevated temperatures, but at a slower pace than the PM1 modes. As a result, the DOS associated with the PM1 modes moves upwards rapidly and merges with that of the PM2 modes at $T=700$~K, as shown in Fig.~\ref{fig:phonon-new}(c).

To understand the reasons for the large frequency-hardening of the PM1 modes, we plot the PES of one of the degenerate zone-center modes in Fig.~\ref{fig:aprnana}(a). Interestingly, this curve features a relatively flat bottom corresponding to a low harmonic frequency, but exhibits a sharp increase in the energy at larger displacements, such as those at elevated temperatures, indicating strong anharmonicity. Specifically, Taylor expansions of the PES based on polynomials around the equilibrium position show strong quartic anharmonicity. This is consistent with the observed significant hardening of the PM1 modes.

\begin{figure}[htp]
	\centering
	\includegraphics[width = 0.85\linewidth]{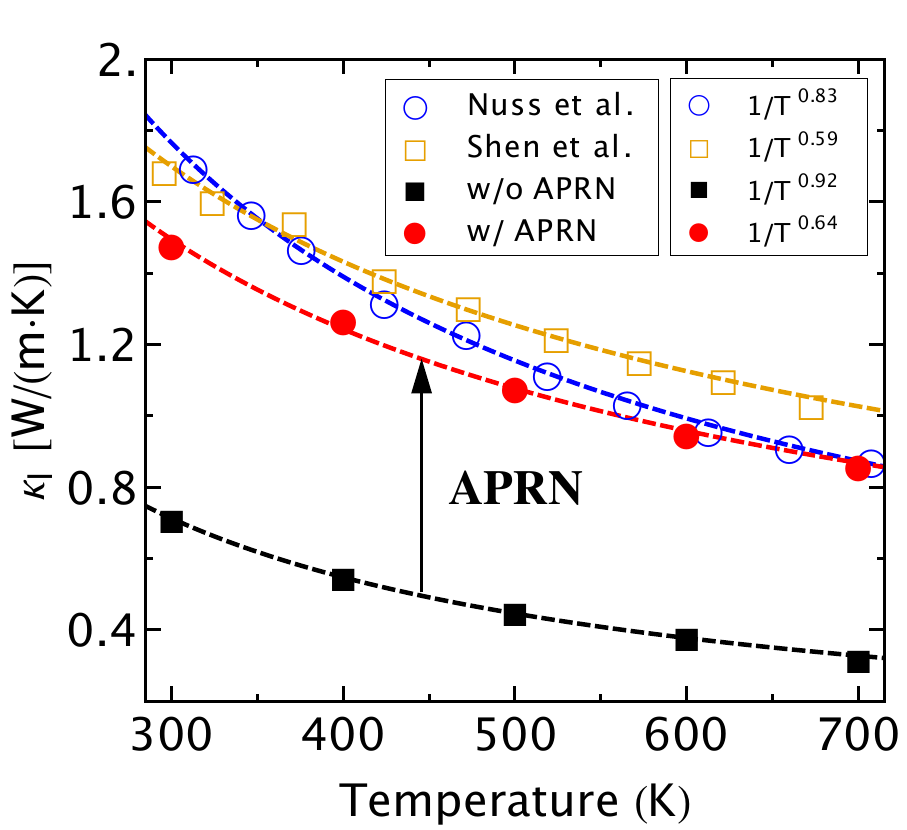}
	\caption{
	Computed lattice thermal conductivities without (filled black squares) and with (filled red circles) anharmonic phonon renormalization (APRN) compared with the experimental measurements by Nuss \etal \cite{Nuss2017} and Shen \etal \cite{Shen2018} The dashed lines represent a fitting for the thermal conductivity in the form of $a/T^b$, where the temperature dependence is shown in the legend.
	 }
	\label{fig:kappa}
\end{figure}

\begin{figure*}[htp]
	\centering
	\includegraphics[width = 1.0\linewidth]{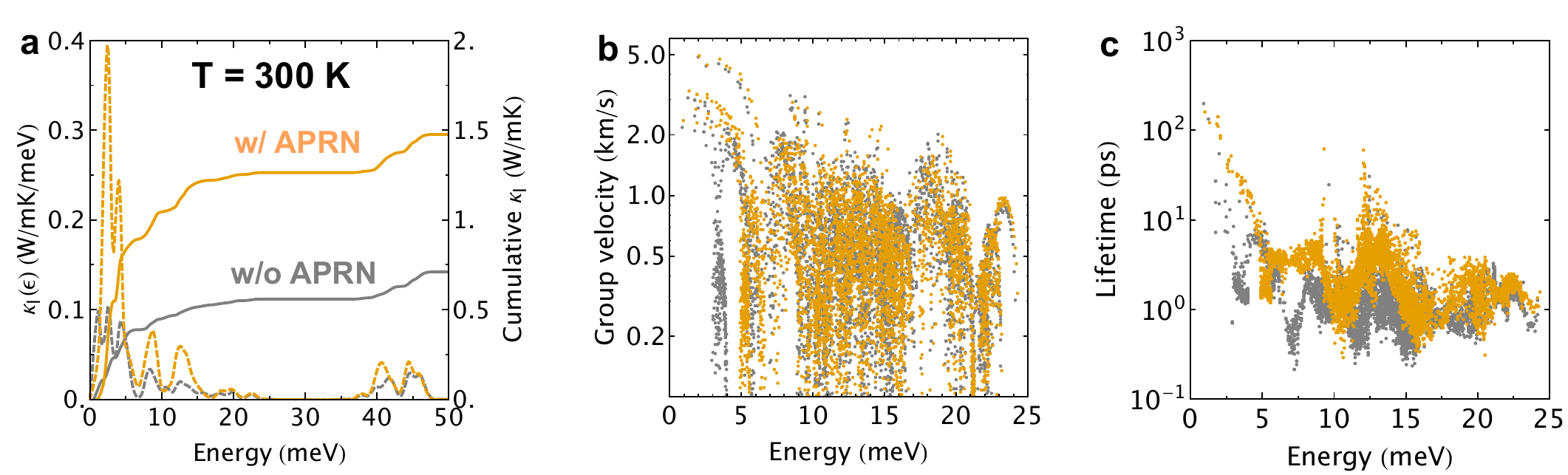}
	\caption{
	Comparison of (a) energy cumulative/differential lattice thermal conductivities, (b) group velocities and (c) lifetimes due to three-phonon interactions calculated without (gray) and with (orange) anharmonic phonon renormalization (APRN) at 300~K.
	 }
	\label{fig:kappa-ana}
\end{figure*}

We next discuss the effects of the anharmonic mode hardening on the experimental data other than the phonon dispersion itself; the latter requires time-consuming and technically complex inelastic neutron scattering techniques on high-quality single-crystals, which may not be practical in many cases. In particular, we consider basic thermodynamic properties that may indirectly reflect the strong temperature dependence of rattling modes, such as the atomic mean square displacements and molar heat capacity; both can be measured with relative ease. Figure~\ref{fig:aprnana}(b) compares the temperature-dependent principle mean square atomic displacements ($U_1$, $U_2$, $U_3$) of the Ag$_{6}$ cluster calculated with and without APRN using the Phonopy software~\cite{Togo20151}. First, we confirm that the displacement parameters of the Ag atoms are considerably larger than those for the other atoms, and that they show a pronounced anisotropy in line with the experimental results of Nuss \etal \cite{Nuss2017}. Second, we find that APRN significantly reduces $U_{1}$ relative to the harmonic value, which renders measuring $U_{1}$ at elevated temperatures an effective way to suggest or confirm the presence of anharmonic mode hardening. Third, in the case of Ag$_6$Ge$_{10}$P$_{12}$ evidence for hardening of the rattling modes can be gained by comparing the theoretical and experimental ratios of the principle mean square atomic displacements $2U_1/(U_2+U_3)$. The calculated $2U_1/(U_2+U_3)$ at $T$ = 300~K with APRN is 2.6, which is in much better agreement with the experimental value of 3.0~\cite{Nuss2017} than the value of 4.0 obtained from the purely harmonic model. Figure~\ref{fig:aprnana}(c) shows the temperature-dependent constant-volume heat capacities calculated using theoretical phonon frequencies at $T=0$~K (harmonic) and $T=300$~K (with APRN), as well as the experimental measurements~\cite{Gmelin1989}. Unsurprisingly, we see that the calculated heat capacity using the harmonic $T$ = 0~K phonon frequencies agrees well with the measured low-temperature ($T <$ 10~K) heat capacity, while using the frequencies of the $T = 300$~K anharmonically renormalized phonons to predict the low-$T$ heat capacity leads to a significant underestimation. This is, of course, because at low temperatures the effect of higher-order anharmonicity is weak and the harmonic approximation is correct. It also shows that, in general, one cannot use the measured high-$T$ phonons to predict the low-$T$ thermophysical properties. In contrast, the heat capacities calculated with and without APRN at $T$ = 300~K are practically indistinguishable. This happens because the contributions to the total heat capacity from the rattling phonon modes are almost saturated to the classical Dulong-Petit limit, and the temperature-dependence of the frequencies $\omega (T) = \omega_0 + aT $ only contributes a term that is proportional to $a^2$. This indicates that the heat capacity data can only be used together with other information to confirm the hardening of rattling phonon modes.

%--------------------------------------------------------------------------------------------------------------------------
\subsection{Lattice thermal conductivity}

Next, we examine the effects of the renormalized phonon frequencies on the lattice thermal conductivity $\kappa_{l}$. Fig.~\ref{fig:kappa} shows the computed $\kappa_{l}$ in comparison with experimental measurements~\cite{Nuss2017,Shen2018}. We see that $\kappa_{l}$ computed without APRN displays significantly smaller values than experimental results. It is worth noting that our calculated $\kappa_{l}$ is also slightly smaller than the earlier theoretical result of Ref.~[\onlinecite{Shen2018}], which is due to the more stringent convergence criteria employed in this study. Intriguingly, $\kappa_{l}$ calculated with APRN exhibits much higher values than without APRN (by a factor of two in the entire temperature range) and achieves a much better agreement with experiments. Interestingly, we find that in addition to increasing the absolute value of $\kappa_{l}$, APRN also strongly alters its temperature dependence. When APRN is small and three-phonon processes dominate, one usually sees an approximately $1/T$ decrease of $\kappa_l$ that is caused by the increase in phonon population with increasing $T$. In marked contrast, strong APRN leads to a noticeably slower decay of $\kappa_{l}$ following a $1/T^{0.64}$ power law instead of the $\kappa_l \propto 1/T^{0.92}$ relation calculated without APRN. This is in rather good agreement with the experimental measurements of Shen~\etal~\cite{Shen2018}, who report $\kappa_{l} \propto 1/T^{0.59}$. However, a faster decay following $1/T^{0.83}$ was found by Nuss~\etal~\cite{Nuss2017}. Such an inconsistency between the two experimental studies may be due to the presence of a possible impurity phase~\cite{Shen2018} and can be resolved by performing measurements on high-quality single crystals. Nevertheless, our results unambiguously confirm that APRN has a significant impact on $\kappa_{l}$ and leads to a remarkably good agreement between theory and experiment.

To gain a better understanding of the impact of APRN on $\kappa_{l}$, we plot the cumulative and differential $\kappa_{l}$ as a function of the phonon frequency in Fig.~\ref{fig:kappa-ana}(a). It is evident that the significant increase in $\kappa_{l}$ mainly comes from the phonon modes with frequencies below 20~meV, and the acoustic modes contribute much more than optical phonons. It is evident from Eq.(\ref{eq:PBTE}) that such a significant enhancement in $\kappa_{l}$ may occur due to APRN-induced changes in the phonon heat capacity, group velocity, or lifetimes. We first rule out the heat capacity since we have found that APRN have negligible effect on heat capacity at 300~K. Furthermore, the effect of APRN  on the group velocity of the acoustic modes is a small increase, while the flat PM1 modes are rigidly shifted upwards and their group velocities remain negligibly low, as explicitly shown in Fig.~\ref{fig:kappa-ana}(b). Hence, change in group velocities cannot explain the two-fold increase of $\kappa_{l}$. We therefore conclude that APRN significantly decreases scattering rates and increases phonon lifetimes. Indeed, comparison of the lifetimes calculated with and without APRN in Fig.~\ref{fig:kappa-ana}(c) shows a significant enhancement for those phonon modes lying below 10~meV, in line with the sharp peaks in the derivative of $\kappa_{l}$ in Fig.~\ref{fig:kappa-ana}(a).

\begin{figure}[h]
	\includegraphics[width = 1.0\linewidth]{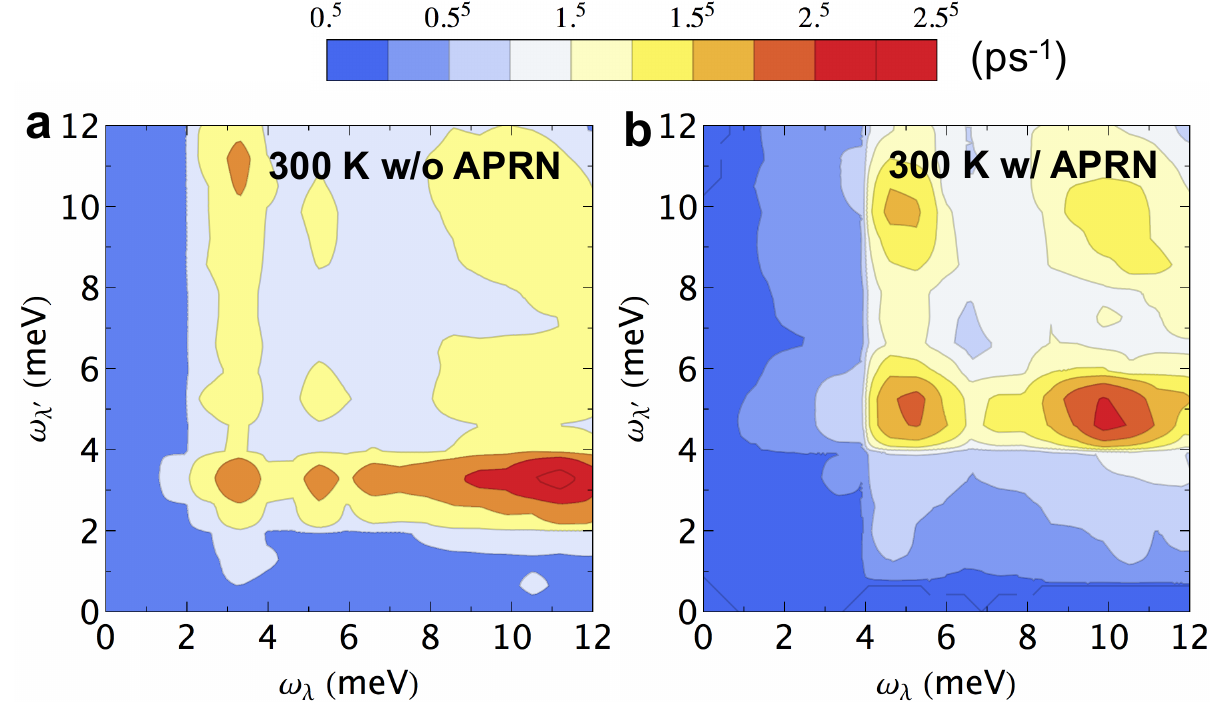}
	\caption{
	Contour plot of the three-phonon combination ($q+q^{\prime} \rightarrow q^{\prime\prime}$) rates at $T=300$~K  calculated with (a) and without (b) APRN.
	 }
	\label{fig:ratesmap}
\end{figure}

To reveal the physical origin of the reduced scattering rates of these low-lying phonon modes, in Fig.~\ref{fig:ratesmap}(a) and (b) we plot the mode-resolved scattering rates associated with the dominant three-phonon combination processes ($q+q^{\prime} \rightarrow q^{\prime\prime}$) as functions of the two phonon frequencies $\omega_{q}$ and $\omega_{q^{\prime}}$. It is evident that the PM1 modes act as effective scattering channels for both acoustic and optical modes. In addition, comparison between Fig.~\ref{fig:ratesmap}(a) and (b) indicates that APRN not only reduces the overall scattering rates but also noticeably alters the scattering landscape. Specifically, in the absence of APRN high scattering rates exist in the low-lying acoustic region (below 5 meV), where we see a strong scattering peak at $\omega_{q}=\omega_{q^{\prime}} \approx$~3.3 meV. In contrast, when APRN is included the scattering peak is shifted to a higher frequency ($\omega_{q}=\omega_{q^{\prime}} \approx$~5.2 meV), which reduces the interactions with and scattering rates for the heat-carrying acoustic modes. A close inspection further reveals that acoustic modes with frequencies in the PM1 range are heavily scattered by interactions with the PM1 modes, highlighting the importance of the proper treatment of frequencies of the ``rattling'' modes for an accurate theory of $\kappa_{l}$ in this and similar compounds.

%\subsection{Remarks}

The current results, despite achieving remarkable agreement with experiment, should be taken with some caution, since such a good agreement might be partly due to a cancellation of errors. An example of a system where such error cancellation occurs is PbTe. It has been demonstrated by one of us that the approximately $1/T$ decay of $\kappa_{l}$ in PbTe stems from a strong interplay among thermal expansion, APRN, three- and four-phonon scattering processes, despite the fact that a much simplified model employing only three-phonon scattering also leads to a $1/T$ decay of $\kappa_{l}$~\cite{yixia2018}. Further refinement of the lattice thermal transport model for Ag$_{6}$Ge$_{10}$P$_{12}$ can be achieved by considering two additional factors. First, inclusion of four-phonon scattering rates may become increasingly important above the Debye temperature, as recently pointed out by several authors~\cite{Tianli2016,Tianli2017,yixia2018,Ravichandran2018,yixiagete2018}. Second, the adopted theoretical framework based on the phonon gas model excludes the nondiagonal Peierls contribution~\cite{Allen1993disorder,Auerbach1984}, which could be important in complex crystals~\cite{Simoncelli2019,Mukhopadhyay1455}. Recent theoretical developments have made it possible to explicitly estimate both contributions~\cite{Tianli2016,Simoncelli2019}, which might be a focus of future investigations. 

%--------------------------------------------------------------------------------------------------------------------------
% Conclusion
%--------------------------------------------------------------------------------------------------------------------------
\section{\label{sec:conclusion} Conclusion}
In summary, we have performed a first-principles study of lattice dynamics and thermal transport properties of Ag$_{6}$Ge$_{10}$P$_{12}$. We go beyond harmonic approximation in computing phonon dispersions by including temperature-induced anharmonic phonon renormalization based on self-consistent phonon theory. Inspections of lattice dynamics reveals strong temperature dependence of rattling vibrational modes associated with Ag atoms, which stems from a potential energy surface with significant high-order contributions. We find that the temperature-induced hardening of rattling modes tends to reduce scattering rates of low-lying acoustic modes and therefore significantly enhance lattice thermal conductivity, improving the agreement between theory and experiment. Limitations and possible sources of error cancellations of the current approach are also discussed.

\textbf{Acknowledgements.}
VO acknowledges financial support from the National Science Foundation Grant DMR-1611507. This research used resources of the National Energy Research Scientific Computing Center, a DOE Office of Science User Facility supported by the Office of Science of the U.S. Department of Energy under Contract No. DE-AC02-05CH11231.

\bibliography{AgGeP}

%\clearpage
%\newpage
\appendix
%\numberwithin{equation}{section}
%\renewcommand{\thesection}{\Alph{section}}
%-----------------------------------------------------------------------------------------
\section{Compare renormalized phonon dispersions using different anharmonic phonon renormalization schemes} \label{sec:compaprn}

\begin{figure}[htp]
	\includegraphics[width = 0.85\linewidth]{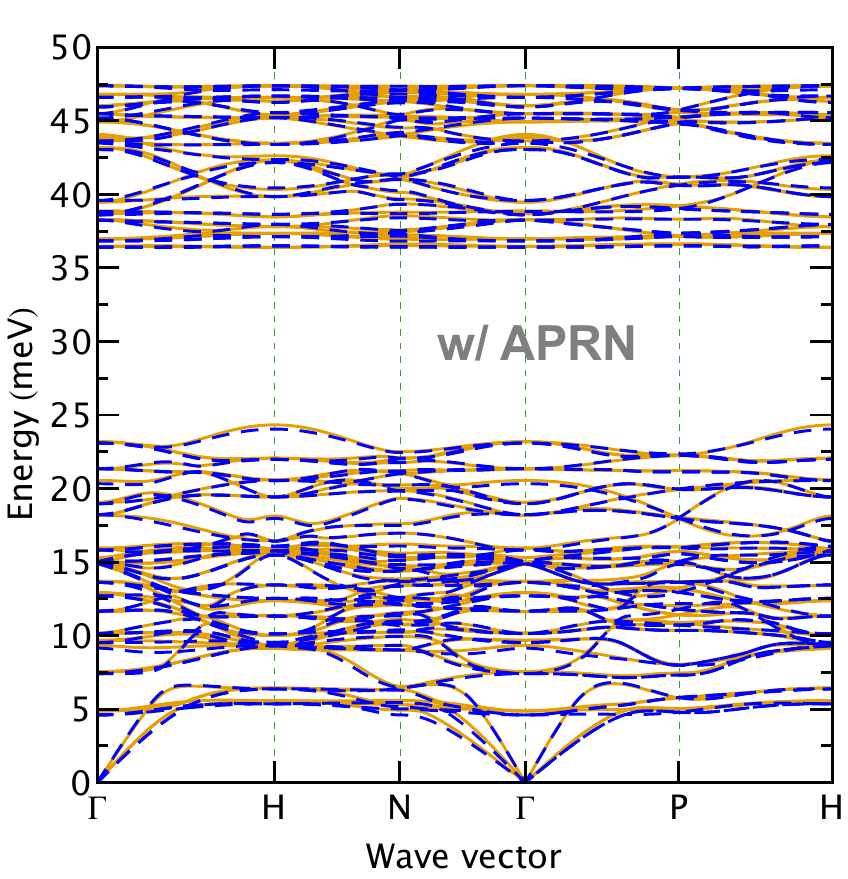}
	\caption{
	Calculated phonon dispersions at 300~K considering APRN. Dispersions colored in orange were obtained using self-consistent phonon theory formulated in the reciprocal space. Dispersions colored in blue were obtained using a real-space-based APRN scheme, as detailed in Ref.~[\onlinecite{yixia2018}] and Ref.~[\onlinecite{yixiagete2018}].
	 }
	\label{fig:compaprn}
\end{figure}

\end{document}